
\magnification \magstep1
\baselineskip 20pt
\def\prref{\par\noindent\hangindent=0.3cm\hangafter=1}

\def \cntl{\centerline}
\def \etal {{\it et al.}}
\def \eg {{\it e.g.,}}
\def \cf {{\it cf.,}}
\def \vs {{\it vs.}}
\def \ie {{\it i.e.,} }

\def \gammao {\gamma_0}
\def \Nature {{\it Nature}}
\def \ApJ {{\it Ap. J.}}
\def \apj {{\it Ap. J.}}
\def \ApJS {{\it Ap. J. Suppl.}}
\def \ApJLet {{\it Ap. J. Lett.}}

\def \apjl {{\it Ap. J. Lett.}}
\def \MNRAS  {{\it M.N.R.A.S.}}

\def \vjec {\vfill\eject}

\def \tablerule{\noalign{\hrule}}
\def \rp{ r^\prime}
\def \r0p{ r{_0^\prime}}

\hfill\ Preprint Number CfPA-95-TH-25
\vskip 0 true cm
\vskip 1. true cm
\cntl{\bf SECONDARY INFALL:  THEORY {\it VERSUS} SIMULATIONS}
\vskip 8pt
\cntl{\bf  S. Zaroubi$^{1,2}$\footnote{}
{\noindent saleem@pac1.berkeley.edu}, A. Naim$^3$ \footnote{}
{\noindent hn@mail.ast.cam.ac.uk}and Y. Hoffman$^2$\footnote{}
{\noindent hoffman@vms.huji.ac.il}}
\medskip
\cntl{(1) Astronomy Department and Center for Particle Astrophysics}
\cntl{University of California, Berkeley, CA 94720, U.S.A.}
\medskip
\cntl{(2) Racah Institute of Physics, The Hebrew University, Jerusalem 91904,
Israel}
\medskip
\cntl{(3) Institute of Astronomy, Cambridge University, Madingley Road,
Cambridge CB3 0HA, UK}

\cntl{\bf ABSTRACT}
\vskip 8pt

The applicability of the highly idealized secondary
infall model to `realistic' initial conditions is investigated. The collapse of
proto-halos seeded by $3\sigma$ density perturbations to an Einstein--de Sitter
universe
is studied here
for a variety of scale-free power spectra with spectral indices
ranging from $n=1$ to $-2$. Initial conditions are set by the constrained
realization algorithm and the dynamical evolution is calculated both
analytically and numerically. The analytic calculation is based on the simple
secondary infall model where spherical symmetry is assumed. A full numerical
simulation is performed by a Tree N-body code where no symmetry is assumed. A
hybrid calculation has been performed by using a monopole term code, where no
symmetry is imposed on the particles but the force is approximated by the
monopole term only. The main purpose of using such code is to suppress
off-center mergers. In all cases studied here the rotation curves calculated
by the two numerical codes are in  agreement over most of the mass of the
halos,
excluding the very inner region, and  these are compared with the analytically
calculated
ones.

The main result obtained here, reinforces the foundings of many N-body
experements, is that the collapse proceeds 'gently' and not {\it via} violent
relaxation. There is a strong correlation of the final energy  of
individual particles with the initial
one. In particular we find a preservation of the
ranking of particles according to their binding energy.  In cases where the
analytic model predicts non-increasing rotation curves its predictions are
confirmed by the simulations. Otherwise, sensitive dependence on initial
conditions is found and the analytic model fails completely. In the
cosmological context power spectra with $n\geq -1$ yields (in the mean)
non-increasing rotation curves, and in such cases the secondary infall model
is expected to be a useful tool in calculating the final virialized structure
of collapsing halos.

\footline= {\hss\tenrm\folio\hss}

\bigskip
\cntl {\bf I. INTRODUCTION}

In a cosmological model in which structure forms via gravitational instability,
the collapse of proto-structures  onto  local  density maxima of the
primordial perturbation field plays a major role in structure formation.
Although the final (non-linear) density maxima do not necessarily emerge
out from initial (linear) density maxima (\cf\   Hoffman 1986, and 1989,
Katz \etal 1993, Zaroubi and Hoffman 1993b and Bertschinger \& Jain 1994)
there is a good evidence to believe a strong connection between initial
high maxima and final density maxima (van der Weygaert \& Babul 1994).

The problem of collapse seeded by such maxima has been investigated from two
different points of view. One is the statistical distribution of the objects
formed this way, namely the distribution of galaxies and clusters. This is
related to the so-called biasing problem which has been extensively studied
(Kaiser, 1984, Davis \etal\  1985 Bardeen \etal\  1986 hereafter BBKS).
The other aspect of the collapse seeded by local peaks is the internal
structure
of
the objects thus formed and its dependence on the statistical properties of the
primordial perturbation field. This problem had been  first addressed by Gott
\&
Gunn (1972) and was followed by the seminal paper of Gunn (1977, hereafter
G77).
The ideas and formalism suggested by G77 have led to   considerable  efforts to
understand the dynamical and statistical aspects of the problem, using both
analytical calculations and numerical N-body simulations to understand this
highly
non-linear collapse process. It is the aim of the present paper to present
N-body
simulations which have been designed to check and make detailed comparison with
the predictions of the analytical calculations.

Gunn \& Gott (1972) studied the very idealized problem of the collapse onto a
single  $\delta_{\rm Dirac}$-like density perturbation to an otherwise
unperturbed
flat Friedmann  universe. This process has been described later on as the
secondary infall, namely the collapse of less bound shells onto a perturbation
that has already collapsed  and virialized. By following the trajectories of
spherical shells, and ignoring the process of shell crossing, the final
virialized structure was calculated. The major breakthrough was done by G77 who
realized that the spherical collapse onto the peak admits an adiabatic
invariant. The case of a self-similar perturbation embedded in a flat
Einstein--de Sitter universe  allows an analytical calculation of the
final virialized structure which takes into account shell crossing. The
pioneering work of G77 was extended by Filmore and Goldreich (1984, hereafter
FG) and Bertschinger (1985). These authors presented a new numerical approach
to the problem of self-similar collapse in a flat universe. These self-similar
solutions provide an exact description of the collapse all the way to the final
virial equilibrium. FG further extended the analytical approach of G77 and
presented an asymptotic analytical solution of the density structure, the
nature of which is described below. A similar analytical calculation was
presented also by Bertschinger and Watts (1988).

The problem of collapse and in particular the formation of dark halos,
hereafter this term describes  bound objects made of collisionless  matter,
was studied in great detail by Quinn, Salmon and Zurek (1986; QSZ), Zurek,
Quinn and Salmon (1988), Frenk \etal\  (1988) and Crone, Evrard and Richstone
(1994).  These authors addressed the
problem by means of N-body simulations, in which the full non-linear evolution
from
typical initial conditions corresponding to different models was followed. The
main
emphasis of these papers is on the dynamical resolution of forming objects and
the
study of their structure. Thus, unlike the analytical approach  where very
idealized
physical systems have been considered, the N-body simulations enable the study
of
the evolution of typical objects whose structure is not self-similar and does
not obey a high symmetry. These studies focused on finding systematic trends in
the forming structures, and in particular possible dependence on the initial
conditions.

In the standard model of cosmology structure emerges out of a primordial
perturbation field. This field is assumed to constitute a random Gaussian
field, whose statistical properties are defined by its power spectrum.  A first
step to bridge the gap between the highly idealized analytical calculations and
the numerical simulations was taken by Hoffman and Shaham (1985, hereafter HS),
who calculated the ensemble mean density profile around high local density
maxima. Assuming that structure around peaks of the primordial field is indeed
given by the ensemble average, and that the dynamics is well approximated by
the
G77
dynamical model, HS calculated the virialized final structure expected in a
variety of cosmological models. The analytical calculations of HS present a
highly simplified picture of the extremely asymmetrical and very non-linear
actual physical process. In spite of this high level of simplification, the
predictions made by HS have been basically confirmed by  N-body simulations
(QSZ, Frenk \etal\  1988). Hoffman (1988) made a
detailed comparison of analytical predictions of the so-called secondary infall
model (hereafter SIM) with the simulations of QSZ and Quinn and Zurek
(1988) and found a  good agreement in terms of the rotation curves of
dark halos that are formed in the simulations. Thus, the simple model
successfully reproduces the virialized structure of halos. However, a closer
inspection of the formation process of these halos reveals a picture that
stands in complete disagreement with the SIM. Halos do not
form by the infall of spherical shells, but rather by the coalescence and
mergers of substructures (QSZ). Nothing of the symmetric and ordered process
envisaged by HS is actually confirmed by the numerical simulations, in which
the collapse process of halos seems to be an extremely unordered and random
process. In spite of these two very different formation pictures, the analytic
and numerical calculations yield a similar final structure. It is this
seemingly paradoxical behavior that the present {\it paper} addresses, by
designing and running N-body calculations so as to probe the basic ingredients
of the collapse process of halos. The main aim of the {\it paper} is to find
the key to the success of the SIM, and at the same time to
determine the limitations of this simple model. A major goal  here is to
define under what conditions the model can be applied to realistic initial
conditions and to serve as a practical model for bound structure formation in
an
expanding universe. An entirely different approach to the problem has been
adopted by Quinn and Zurek (1988) and Warren \etal\  (1991), who suggested that
the basic mechanism is related to the mergers of subclumps rather than
centralized collapse. The role of mergers will be studied here by the N-body
simulations.
 The structure of the {\it paper} is as follows. In \S II the main
features of the analytical model are presented and the basic question which
are to be solved by the numerical simulations are posed. This is followed by
the description of the numerical simulations, in \S III, and the initial
conditions, in \S IV.
The results of the N-body simulations are analyzed in \S V. The {\it paper}
concludes with a discussion given in \S VI.

\bigskip
\cntl{\bf II. Secondary Infall}

The main essence of the SIM is that for a single spherical
scale-free perturbation to an otherwise homogeneous flat  universe, the final
time averaged radius of a given Lagrangian shell scales   with its initial
radius. Thus, for a primordial power law density perturbations to an
Einstein--de
Sitter universe one expects a self-similar evolution. For the case of a
Gaussian
perturbation field the
mean density profile around high local maxima of the density perturbation field
is given by the autocorrelation function.
In the case of a scale free field
whose power spectrum is given by $P(k)\propto k^n$, where
the relevant exponents in cosmology are in the range of  $-3 < n <  1$, one
finds (HS):
$$ \delta(r_i)\propto r{_i^{-(3+n)}},\eqno(1)$$
(Here $r_i$ is the initial radius defined at some arbitrary time, well within
the
linear regime.) A fiducial density profile is defined  as the structure one
would obtain by freezing all shells at their turn-around radius. This yield
(HS):
$$ \rho_0(r) \propto r^{-\gammao}, \eqno(2)$$
where
$$ \gammao = {3(3+n)\over 4+n}.\eqno(3)$$
Assuming that the final (time averaged) radius $r$ of the
shell labeled by $r_i$  scales with $r_i$, one finds that the final density
run,
$ \rho(r)$ scales with the fiducial one, namely $\rho(r)\propto r^{-\gammao}$.

In the case of a centralized perturbation, whose inner part is denser than its
outer regions, one expects a secondary infall, namely the inner denser part
collapses and relaxes first and the outer shells are collapsing on a longer
time scale. As was pointed first by G77,   the dynamics of the
infalling less bound shells   admits the product $r M(r) $ as an adiabatic
invariant, where $M(r)$ is the mass that is momentarily interior to the radius
$r$. Expressing the scaling of the final radius with the initial one by
relating $r_f$ to the turn around  radius $r_0$ we write $r_f = F r_0$, where
for a
self similar infall $F$ is a constant that depends on $\gammao$. Now,
the one-to-one dependence of $r_0$  on $r_i$ is easily found from energy
considerations. Given all that, the adiabatic invariance implies that the
rotation curve of the halos is given by:
$$
V^2_{rot} ={ GM_f(r_f)\over r_f} = {GM_0(r_0)\over F^2 r_0}\eqno(4)
$$
Here $M_0(r_0)$ is the mass enclosed by the shell labeled $r_i$ at the turn
around epoch. To complete this analytical calculation one should find the
collapse, or contraction, factor $F$.

For spherical systems, the mass interior to a given, \ie\ Lagrangian shell, is
conserved up to its turn around. However, beyond this stage shell crossing
occurs  and this mass is not conserved any longer.
The equation of motion of spherical shells admits a self-similar solution which
transforms the partial differential equation into an ordinary one, which is
solved numerically (FG, Bertschinger 1985). These numerical exact solutions can
be obtained analytically in the asymptotic limit of a shell which has past its
turn around phase much earlier than some given time, defined as
the `present' time, and whose  apocenter is much smaller than the current turn
around
radius, $R_0$. These asymptotic solutions have been solved subject to
the approximations of spherical symmetry, scale invariance and the validity of
adiabatic invariance (G77, FG, Bertschinger and Watts 1988). This problem has
been
recently addressed and an
exact analytical solution for the collapse factor $F$ in the asymptotic limit
has been obtained (Zaroubi and Hoffman 1993, hereafter ZH).
This
solution is presented here and used in the comparison of analytical and
numerical calculations. Consider a shell
whose current apocenter (which is approximately equal to the time-averaged
radius)
is $r$ and whose turn around radius is $r_0$, then the collapse factor is given
by:
$$
F=\left(1+(3-\gammao)\int_{1}^{U} u^{2-\gammao} P({1\over u})
du\right)^{-1}\eqno(5)
$$
Here $P(r/r')$ is the fraction of time a particle with apocenter $r'$ spends
inside $r$, $U=R_0/r_0$, $R_0$ is the outer radius of the system or the present
epoch turn-around radius and $r_0$ is the turn-around radius of the shell whose
current apocenter is $r$. The crucial step is the evaluation
of $P(u)$. Various  approximations were taken by   G77,  FG and
Bertschinger (1985) and it has been solved exactly by ZH. Now, the integral in
the denominator of Eq. 5 shows a very different behavior
for $\gammao$ smaller or larger than $2$ (which corresponds to $n=-1$, in
the limit of self similar systems). In the limit of $U\gg 1$ the
integral is dominated by its lower limit for $\gammao > 2$ and by its upper
limit for $\gammao < 2$. Thus for $\gammao > 2$ the dynamics of the shell
labeled
by $r$ is affected mainly by nearby outer shells such that $\rp {^>_\sim}r$,
while in the other case of $\gammao < 2$ it is dominated by shells that have
turned around only recently, $\r0p{^<_\sim}R_0$ and $\r0p$ is the turn around
radius corresponding to $\rp$. It follows that in the above limit one
finds self-similarity for $\gammao>2$. In the other case of
$\gammao \leq 2$ the final density run is a power law of $2$ . Note, that in
the
case of $\gammao>2$ the effect of shell crossing is a local one, as only nearby
shells of
$\rp {^>_\sim}r$ contributes to the mass interior to $r$. Therefore, in such a
case one
might hope that the model can be applied over a finite range of $r$, even if
globally the
structure does not follow a power law and self-similarity.  For
$\gammao\leq 2$ the main contribution to $M(r)$ comes from the outer shells
that
have
turned around only recently. This introduces a very sensitive dependence on
boundary
conditions. The appropriate upper limit for the integral of
Eq. 5 is $U\approx 3$ (ZH). For such a value the dependence of $F$ on $\gammao$
is
fitted by: $$ F=0.186 + 0.156 \gammao + 0.013 \gammao^2+ 0.017\gammao^3-0.0045
\gammao^4+ 0.0032\gammao^5 \eqno(6) $$

The analytical model presented here relies on three basic assumptions, namely
spherical symmetry, scale invariance and the validity of the adiabatic
invariance. Yet, for actual halos emerging out of Gaussian random fields it is
clear that at least the first two assumptions do not apply. The structure
around typical peaks is neither spherical nor is it characterized by a power
law (BBKS) and it is certainly finite. The question thus arises
is to what extent can the simple model describe the complex dynamical
evolution. Now, from previous numerical simulations  we know that at least in a
statistical
sense the model does correctly yield the final structure of bound halos
(QSZ, Frenk \etal\  1988;
for detailed comparison see Hoffman, 1988). Thus, applying the model to
ensemble averaged density profile reproduces the ensemble average
taken over all the objects in the simulations, the dynamics of each one of
which is properly calculated by the N-body codes.

\bigskip
\cntl{\bf III. Numerical Simulations}
We model here the collapse of bound structures seeded by local density maxima
of
the
primordial perturbation field in a flat Friedmann universe. The perturbation
field is assumed to be a Gaussian random field which is defined by its power
spectrum. For a scale free process  the power spectrum is given by $P(k)\propto
k^n$. Given a particular realization of the perturbation field, constrained to
be
centered on a local density  maximum, the subsequent dynamical evolution is
calculated by three different methods.

First, the final virial configuration is calculated by using the SIM. This is
done by taking the spherically averaged $\delta$-field (centered on the peak),
and applying Eq. 4 to it. Thus, the analytical model that has been derived for
scale-free initial conditions (ZH), is applied here to a general initial
density profile which is not self-similar. The collapse factor $F$ is
calculated here for each given spherical shell of radius $r$ by using the
local logarithmic derivative of the initial density profile. The self-similar
solutions are applied here locally, as if the shell crossing process depends
on the local properties of the density field. In the case where the actual
density profile yields a  fiducial density which follows an $r^{-\gammao}$
power law with $\gammao>2$, the actual $\gammao$ thus obtained is used to
calculate the collapse factor $F$. However,  for $\gammao<2$ we take it to be
$\gammao=2$. The dependence of $F$ on $\gammao$ is calculated by ZH. This
relation has one free parameter, namely the ratio of the largest turn-around
radius (at a given time)  to the turn-around radius of the shell under
consideration. Based on the N-body simulations here we estimate this ratio
to be $U=3$ in average.

Second, the final virialized halo is calculated by evolving the initial system
in full N-Body simulations. These N-body simulations are carried out by
using the Treecode algorithm (Barnes and Hut, 1986, and Hernquist
1987, 1988, 1990).
Third, we introduce a new numerical model to trace the evolution of high
density
peaks under the assumption of spherically symmetric force and realistic initial
conditions, rather than the spherically symmetric force and initial conditions
as
assumed in the SIM. This code is used to trace the
evolution of a single peak in an otherwise Einstein--de Sitter universe. Such a
code stands in  between the full N-body calculation and the simplified
analytical
model. It turns out that the rotation curves of the virialized systems
calculated using this method or with the Treecode, agrees   well over most of
the
halos, excluding the very inner regions (see $\S V$). Therefore  this method is
used here
as a reference for testing the SIM predictions for most of the simulations.

The location of the center of the peak  is very crucial for this kind of
approach and it should be carefully defined; here we chose it to
coincide with the point of maximal density. The
time step used in this code is taken to be a fixed fraction from the dynamical
time ($\tau_{dyn}$) of the system. Where $dt= 10^{-4} \tau_{dyn}$ gives a very
good energy conservation (${\Delta E \over E}\leq 1 \%$) over a complete
simulation. A softening parameter is used to avoid an infinite force at the
origin. The simulated region has been chosen to be a sphere of radius
$R=1$, which along with fixing the gravitational constant to unity
($G=1$) define the physical units of the simulation. The softening
parameter of most of the simulations is $0.1$.

The two basic limitations of the present simulations are the nature of boundary
conditions and the rather small number of particles. Here vacuum boundary
conditions are assumed and therefore tidal interactions are ignored. As for the
number of particles, in most runs $N=2000$ particles are used. For one of the
runs this has been increased to $N=4000$ and $8000$, in such a way that the
increase in the number of particles was used to increase the extent of the
simulated objects and not the code resolution. Now, these limitations certainly
affect the outcome of the simulations. However, given that the simple model has
already been successfully confirmed by   large scale simulations (\eg\
QSZ), in which boundary conditions are properly handled,
and given that our simulations are aimed at
studying
in a controlled way the dynamics  of individual objects, we think that the
present calculations are adequate for addressing the problems stated here.

\bigskip
\cntl{\bf IV. Initial Conditions}

The formalism of constrained realizations of Gaussian random fields (Hoffman \&
Ribak
1991) has been used to set the initial conditions.
Here we are interested in making a particular realization of a Gaussian
perturbation field, which is constrained to have a
$3\sigma$ density peak located at the center of the computational box.
A peak is specified by $10$ constraints, namely  the peak amplitude, the three
first
partial derivatives which are set to zero, the three eigenvalues of the
second-order derivatives which form a $3\times3$ symmetric
matrix, and the three angles which
define the direction of the eigenvectors. The three eigenvalues which define
the shape of the peak  are taken to have
their mean values (BBKS), and the eigenvectors are taken along the
Cartesian axis of the computational box. The  realizations are designed to
have such peaks at the center, but they have also other unconstrained local
maxima and minima.

The physical systems  simulated here are self-similar in the mean. The
background
universe is Einstein--de Sitter and the power spectra considered here are scale
free,
$ P(k) \propto k^n  \; \{ n = -2,-1,0,1 \} $.
Two sets of random numbers, which specify the phases and amplitudes and hence
the given realization, are
constructed here and are used for the different power spectra considered.
The two realization are labeled as {\it a} and {\it b}.
 The number of particles used for
each simulation is typically $\approx 2000$ except in two cases where we used
$4000$ and $8000$ particles with the hybrid code to test the effect of larger
systems.

A summary of the  numerical experiments calculated here is given in Table
I. The simulations vary with respect to power spectrum ($n$), family {\it a} or
{\it b} of realizations, number of particles, and the kind of code used. All
models were calculated with the hybrid (symmetrical) code, some of which were
calculated also by the full Tree-N-body code.

\cntl{\bf [ Table 1 ]}

\bigskip

\cntl {\bf  V. RESULTS }
\bigskip

Here we are interested in the gross density structure of the forming halos and
this  is
described by the rotation curves, thus ignoring much of the fine structure.
The resulting rotation curves calculated in the
three different ways are given in Figs. 1(a-c) of three
models of family b with  power spectra of $n=-2,-1,0$. Other models for which
the
rotation curves are calculated by the (analytic) secondary infall
and the hybrid numerical algorithms are given in Fig. 2(a-c).
A close inspection of the calculated rotation curves shows, first, the good
agreement between the full N-body and hybrid codes over most of the mass of the
halos.
There is a discrepancy at the very inner region of the systems. The
comparison of the analytically expected and numerically calculated curves is
more difficult
to interpret. In cases where the local effective $\gammao$ is smaller than $2$,
the
rotation curve is predicted to increase with radius,  corresponding to a
diverging mass
profile, and the   structure is determined by the latest infalling particles.
Not surprisingly the numerical simulations cannot reproduce such
rising rotation curves; this has also been confirmed by higher resolution
N-body
experements of QSZ and Crone \etal (1994). However, for models with spectral
index larger than $-1$ the
effective $\gammao$ increases and a better agreement is expected. Indeed, in
the
case of
the two realizations ($a,b$) of $n=0$ and in particular $n=1$ the predicted
rotation
curves are close to the simulated one. Yet, in all cases the numerically
calculated curves
fall below the analytical ones, indicating a larger collapse factor ($F$) than
the
calculated factor.

The N-body simulations of halos with predicted rising rotation curves poses
severe numerical problems because of its extreme sensitive dependence on the
boundary
conditions. Previous simulations of a representative computational box that
includes many halos and an $n\leq -2$ power spectrum yielded non increasing
rotation curves (\eg\ QSZ). To  study this problem the $n=-2$ model ($b$)
has been further simulated with an increasing number of particles, $N=2000,
4000$ and $8000$, using the hybrid code.
The added particles are used to simulate a larger initial volume and
not to increase the dynamical resolution. The outcome is presented in Fig. 3,
where
the calculated rotation curves are presented and compared with the expected
curve for the $N=8000$ simulation. The change of the simulated curves
with N is expected. In the $N=8000$ case we find a flat rotation curve over
about the inner  half  mass that is followed by a decreasing curve. Note that
the
predicted curve is now dominated by the outer shells and is decreasing.
This manifests the sensitive dependence on the boundary conditions and
shows that the SIM fails in the $\gammao\leq 2$ case. The
model is validated, however, in the $\gammao > 2$ regime of declining curves.

A possible explanation of this surprising agreement between the analytic model
and the
simulations lies in the ordered behavior of the collapsing systems in energy
space as
opposed to the  random behavior in real (configuration) space (Hoffman, 1988).
As was noticed by Quinn and Zurek (1988) the seemingly very complicated
collapse
process looks very ordered and 'gentle' when viewed in energy space. This leads
us to the
key question to be addressed here, namely to what extent a collapsing system
'remembers' its
initial conditions.   In particular we focus here on the energy (per unit mass)
behavior of
individual particles, which changes in time as $(d/dt)(\phi
+v^2 /2)=(\partial/\partial t)\phi$. Therefore in collapsing systems the energy
of
particles changes in time, however it is not clear {\it a priori} whether this
change is
coherent  or leads to a substantial phase mixing. This is studied here by
drawing scatter
plots of the final \vs\  the initial energies of individual particles in all
the
hybrid-code
simulations (Figs. 4 (a-f)). The full Tree-N-body simulations yield similar
scatter
plots. A strong correlation exists in all cases, indicating a coherent and
ordered
evolution in energy space. It is clear that such systems do not go through an
efficient
phase mixing, and thus they retain the memory of their initial conditions. Note
that
there is a clear connection between this correlation and the spectral index
$n$,
the
tightest correlation corresponds to the $n=1$ case and it decreases with $n$.
This trend
coincides with the dependence of the agreement between the  SIM and the
numerical
simulations on the spectral index.

To further study the energy evolution we rank the particles according to their
binding
energies, from the least to the most bound particle.  The possible change of
this ranking
throughout the collapse and virialization process is studied by plotting
histograms of the
frequency of the relative change of this ranking of all particles of a
given hybrid-code simulation Figs. 5(a-g).  One
finds that in  all cases some $70\%$ of particles do not change their rank by
more than
$10\%$, and a very similar result is found in the full Tree-N-body simulations.
The approximate conservation of the
ranking throughout the dynamical evolution  proves that the cosmological
collapse
proceeds rather 'gently' and involves no violent relaxation.

Next, the trajectories of individual particles are considered. In Figs. 6(a-b)
the
trajectories in the $(r-t)$ plane  of 20 randomly selected particles from the
(-1;b) model
are presented. Note, that this is the most successful model in terms of
reproducing the
numerically calculated rotation curve. Yet, these trajectories are in complete
disagreement with the ones calculated by FG and envisaged by the simple
analytic
model.
Very few trajectories do show the gradual shrinking of the turn-around radius.
In the light of this disagreement, the success of SIM in predicting the
the correct rotation curves seems to be very surprising. However one can
argue that the rotation curves depend mainly on the energy distribution,
which is much more smoother and more rounded than the underlying density
distribution, that can account for the agreement between the model and the
simulations (see \S VI for more detailed discussion).

\bigskip
\cntl { \bf VI. DISCUSSION}

The SIM has been rigorously
formulated for self similar systems, yet in the canonical cosmological model
structure arises from a random perturbation field and proto-structure are
necessarily finite and not self-similar. The question that is then naturally
asked is to what extent this highly symmetric model is applicable to
`realistic' initial conditions.  Previous numerical  work (\eg\  QSZ) has
already
proved the validity of the model when applied to N-body simulations and that in
general the simple model correctly predicts the structure of bound objects, at
least in the statistical sense. (This holds in the $n\ge -1$ case.) Yet, closer
inspection of these simulations shows that the actual collapse process seems to
be in complete disagreement with the one predicted by the simple model. It is
this ambivalent behavior that has been studied here. Using the algorithm of
constrained realizations special care has been given to the setting of the
initial conditions of the simulated objects. Given the initial configurations,
these were evolved dynamically to virial equilibrium in three different ways,
namely   by   `exact' N-body simulations, the analytical model and the hybrid
N-body `monopole term' code. Thus, a given initial configuration has been
evolved by three different methods and the final results are compared. Our
basic conclusion is that within the limitation of the numerical simulations the
gross agreement between the model and the large scale N-body simulation is
confirmed at the level of individual objects, in cases where such agreement is
expected.

Our main motivation here is to perform controlled numerical `experiments' of
the collapse onto local density maxima, with an emphasis on the controlled
aspect. This is achieved here in two ways. One is the use of constrained
realizations for setting the initial conditions. The other is the use of the
hybrid monopole term code that stands between the highly symmetric and
analytic SIM and the full N-body simulation. The comparison of the
outcome of this code with that of the standard Tree N-body code allows us to
separate the dynamical effects from that of (the lack of) spatial symmetry. We
find here that within the technical limitation of our simulations, namely the
lack of tidal interactions, the two modes of simulations are in a good
agreement. This proves that mergers mechanism does not stand at the basis of
the success of the SIM.

The basic physical understanding that emerges out of the present collapse
`experiments' is that of two different dynamical regimes depending on the
logarithmic derivative of the (spherically averaged) fractional overdensity as
a function of radius, \ie the effective power law index $n$ or equivalently the
effective $\gammao$ of the fiducial turnaround density profile. In the
cosmological context these regimes correspond to a primordial perturbation
field whose power spectrum is dominated by the high wave number modes $n\ge -1$
or low wave number modes $n\le -1$.
In the regime of highly
correlated perturbation field, $n\le -1$ (\ie in the mean $\gammao<2$), there
is
a strong dependence on the boundary conditions, the dynamics strongly depends
on
the last collapsing shells. In such cases the energy of particles changes
violently in time and one expects to reach at the end a state of statistical
equilibrium, much in the spirit of the Violent Relaxation proposed by
Lynden-Bell
(1967). It is important to note that the two very different approaches, namely
the violent relaxation and the SIM, predict the same final
structure of an asymptotic $r^{-2}$ density profile in the limit of infinite
spherical
system. In the other regime of
$n\ge -1$ the primordial structures are sharply peaked and the dynamics of
already collapsed shells is hardly affected by the ongoing collapse of more
distant and less bound shells. In such a case order is being preserved in
energy space, the initial conditions are well `remembered' by the system and
the final structure reflects the initial conditions in a manner described here.

The regular and ordered evolution in energy space suggests a possible
explanation as to why a model based on spherical symmetry  predicts correctly
the final
mass distribution of collapsing halos. There is a high correlation between the
initial
and final energies of individual particles, and in particular the ranking in
energy
space is roughly preserved. Now, the virial structure depends on, and is
basically
determined by, the final energy distribution, and consequently it depends on
the
initial
energy distribution. In the linear theory of gravitational instability the
total
energy
is determined by the gravitational potential, which is always much smoother and
more
rounded than the underlying mass distribution. Thus, the simple analytic model
which
seems at first glance to (incorrectly) describe the mass distribution, provides
quite a good description of the initial energy distribution of particles, which
in
turn determines the final structure, in the case of $\gammao\le 2$.

The results presented here, and in particular the close agreement with the
prediction of a  spherical top-hat model, seem to be in conflict with earlier
calculations on the role of shear in the gravitational collapse. These show
that
in the quasi-linear regime of gravitational instability in an expanding
universe
the shear (in the velocity field) accelerates the collapse and acts as a source
of gravity (Hoffman 1986 and 1989, Zaroubi and Hoffman 1993b, Bertschinger and
Jain 1994, Eisenstein and Loeb 1995). The shear, which is primarily induced by
the tidal field, represents a deviation from spherical symmetry and thus its
effect seems to be inconsistent with the present results.
Based on this Bertschinger (1994) postulated that local density maxima are not
the
sites where collapse occurs first. This, and the numerical simulations of Katz
\etal\ (1993), shed doubt on the (linear) peaks -- halos association. Yet,
the more recent simulations of van de Weygaert and Babul (1994), which were
designed  to study the effect of shear, have reaffirmed that high
enough (${^>_\sim}2\sigma$) peaks  evolve to form halos. They found, however,
that
shear affects the outer envelopes of halos. The simulations presented here were
not
designed  to study these effects and the limited dynamical range does not allow
proper
modeling of the external shear, however the internal shear which arise from
the aspherical local matter distribution is well presented. This affects
the collapse significantly in the quasi linear regime (Zaroubi and Hoffman,
1993b),
yet the final virial structure is consistent with the top-hat model prediction.
Note
also that in N-body simulations where the computational box contains a typical
patch
of the universe with many proto-objects, both internal and external shears are
properly
sampled. Yet, such simulations basically confirm the predictions of the simple,
top-hat
spherical model (QSZ, Frenk \etal\ 1988, Crone, Evrard and Richstone, 1994).
A tentative conclusion is that shear affects the dynamical evolution in the
quasi-linear regime only, but the virial structure is hardly affected by it.
However,
this should be further tested in controlled N-body simulations.

\bigskip
\cntl {\bf Acknowledgments }

This paper is dedicated to the memory of the late Prof. Jacob Shaham, who
guided

us in our first steps in the study of the secondary infall model and kept his
interet
in the problem ever since.
We would like to thank L. Hernquist for providing us with his
Tree-N-body  code. This work was supported in part by the Israel science
Foundation
grant 590/94, and by the Hebrew University Internal Funds.

\bigskip

\vjec

\cntl{REFERENCES}
\prref Bardeen, J.M., Bond, J.R., Kaiser, N. and Szalay, A.S. 1986, \ApJ, {\bf
304}, 15.
\prref Barnes, J. and Hut, P.1986, \Nature,{\bf 324},446.
\prref Bertschinger, E., 1994, Proc. of the 9$^{th}$ IAP Conference {\it Cosmic
Velocity Fields}, eds. F. Bouchet and M. Lachi\'eze-Rey, (Gif-sur-Yvette Cedex:
Editions Fronti\'eres), pg. 137.
\prref Bertschinger, E. and Jain, B., 1994, \apj, {\bf 431}, 486
\prref Binney, J. and Tremaine, S. 1987, ``Galactic Dynamics", Princeton
University Press, Princeton.
\prref Bouchet, F.R. and Hernquist, L. 1988, \ApJ, {\bf 319}, 575.
\prref Crone, M.M., Evrard, A.E., and Richstone, D.O., 1994, \ApJ, {\bf 434},
402
\prref  Davis, M, Efstathiou, G,  Frenk,C.S.and White, S.D.M. 1985, \ApJ, {\bf
292}, 371.
\prref Eisenstein, D.J., \& Loeb, A., 1995, {\bf 439}, 520
\prref Filmore, J.A. and Goldreich, P. 1984, \ApJ, {\bf 281},1.
\prref Frenk,C.S., White, S.D.M., Davis, M. and Efstathiou, G. 1988, \ApJ, {\bf
351}, 10.
\prref Gunn, J.E., 1977, \ApJ, {\bf 218},592.
\prref Gunn, J.E. and Gott, J.R. 1972, \ApJ, {\bf 176},1.
\prref Hernquist, L. 1987, \ApJS, {\bf 64},715.
\prref Hernquist, L. 1988, {\it Comp. Phys. Comm.}, {\bf 48},107.
\prref Hernquist, L. 1990, {\it Journal of Computational Physics}, {\bf
87},137.
\prref Hoffman, Y. 1988, \ApJ, {\bf 328},489.
\prref Hoffman, Y. and Ribak, E. 1991, \ApJLet, {\bf 380}, L5
\prref Hoffman, Y. and Shaham, J. 1985, \ApJ, {\bf 297},16.
\prref Katz, N., Quinn, T., and Gelb, J.M., 1993, \MNRAS, {\bf 265}, 689.
\prref Lynden-Bell, D. 1967,\MNRAS, {\bf 136}, 101.
\prref Peebles, P.J.E. 1980,``The Large Scale Structure of the Universe",
Princeton University Press, Princeton.
\prref Quinn, P.J., Salmon,J.K. and Zurek, W.H. 1986, \Nature, {\bf 322}, 329.
\prref Quinn, P.J. and Zurek, W.H. 1988, \ApJ, {\bf 331}, 1.
\prref van de Weygaert, R. and Babul A., 1994, \apjl, {\bf 425}, L59
\prref Warren, M.S., Zurek, W.H., Quinn, P.J. and Salmon,J.K. 1991, in the
Proceedings of {\it After the First Three Minutes,} ed. Holt, Trimble
and Bennet.
\prref Zaroubi, S. and Hoffman, Y. 1993a, \ApJ, {\bf 414}, 20.
\prref Zaroubi, S. and Hoffman, Y. 1993b, \ApJ, {\bf 416}, 410.

\vjec

\cntl{\bf TABLE 1}

\bigskip

\bigskip

\vbox{\tabskip=0pt \offinterlineskip
\halign to 343.75pt{\strut#& \vrule#\tabskip=1.em &
\hfil#&\vrule#& \hfil#\hfil& \vrule#& \hfil#\hfil& \vrule#&
\tabskip=1.em minus1em\hfil#\hfil& \vrule#& \hfil#\hfil&
\vrule#& \hfil#\hfil& \vrule#& \hfil#&\vrule#
\tabskip=0pt\cr\tablerule & &\multispan9\hfil  The
Parameters Of The Simulations\hfil&\cr\tablerule & &Label& &Power Spectrum & &
Family & & Particles' \# &&
 Simulations &\cr\tablerule
&&   0;a    && n= 0 && a && 2000 &&  Symm.\rlap* &\cr\tablerule
&&   1;a    && n= 1 && a && 2000 &&  Symm. &\cr\tablerule
&&  -2;b    && n=-2 && b && 2000 && Symm. + Nbody\dag &\cr\tablerule
&&  -1;b    && n=-1 && b && 2000 && Symm. + Nbody &\cr\tablerule
&&   0;b    && n= 0 && b && 2000 &&    Symm.    &\cr\tablerule
&&   1;b    && n= 1 && b && 2000 && Symm. + Nbody &\cr\tablerule
&&  -1;b,4  && n=-1 && b && 4000 &&    Symm.    &\cr\tablerule
&&  -1;b,8  && n=-1 && b && 8000 &&    Symm.    &\cr\tablerule
\noalign{\smallskip}
&\multispan7\dag ~Nbody, indicates full N-body simulation\hfil\cr
&\multispan7* Symm., indicates hybrid code simulation\hfil\cr}}

\vjec

\cntl{\bf Figure Captions}

\item{\bf Figure 1. :}  The final rotation curves calculated in the three
methods, the theoretical model (dashed lines),  the hybrid monople-term code
(long dashed  lines),
and the full N-body code (solid lines). The curves are drawn for
the models labeled (see table I) (a) -2;b, (b) -1;b, and (c) 1;b

\item{\bf Figure 2. :} The final rotation curves calculated   theoretically
(dashed
lines) and numerically by the hybrid code (long dashed lines). The curves are
drawn for the
following models (a) 0;a, (b) 0;b, (c) 1;a.

\item{\bf Figure 3. :} Rotation curves of the $n=-2$ model (labeled -2;b)
calculated
with the hybrid code for N=2000 particles (solid line), N=4000 (dotted line),
and
N=8000 (small dashed line). This is compared with the theoretically expected
rotation
curve for N=8000 (long-dashed line). Note that the length scale here differs
from that
of Fig. 1a.

\item{\bf Figure 4. :} A scatter plot of the final  energies compared with the
initial
ones of the bounded particles as calculated with the hybrid code. The different
graphs
corresponds to the following models (a) 0;a, (b) 1;a, (c) -2;b, (d) -1;b,
(e) 0;b and (f) 1;b. The correlation coefficient is denoted on each graph.

\item{\bf Figure 5. :} Histograms of the change in the ranking (by binding
energy) of the
particles in the  energy  space, as calculated by   the hybrid code. The
histograms
show the percentage of particles which changed their `energy' ranking
within a certain percentage. The models here are the same as in Fig. 4.

\item{\bf Figure 6. :} The radial trajectories of randomly selected particles
tin the Tree-N-body   simulation for the model 1;b. All radii are normalized by
the
maximal turn-around radius of each model.

\bye